\documentclass[twocolumn,english]{revtex4-1}
\usepackage[T1]{fontenc}
\usepackage[latin9]{inputenc}
\usepackage{amstext}
\usepackage{graphicx}
\usepackage{esint}
\usepackage{babel}
\begin{document}

\title{Contractile and chiral activities co-determine the helicity of swimming droplet trajectories}

\author{Elsen Tjhung}
\affiliation{Department of Applied Mathematics and Theoretical Physics, University of Cambridge, Wilberforce Road, Cambridge CB3 0WA, United Kingdom.}

\author{Michael E. Cates}
\affiliation{Department of Applied Mathematics and Theoretical Physics, University of Cambridge, Wilberforce Road, Cambridge CB3 0WA, United Kingdom.}

\author{D. Marenduzzo}
\affiliation{SUPA, School of Physics and Astronomy, University of Edinburgh, JCMB Kings Buildings, Peter Guthrie Tait Road, Edinburgh EH9 3FD, United Kingdom.}

\begin{abstract}
Active fluids are a class of non-equilibrium systems where energy is injected into the system continuously by the constituent particles themselves.
Many examples, such as bacterial suspensions and actomyosin networks, are intrinsically chiral at a local scale, 
so that their activity involves torque dipoles alongside the force dipoles usually considered.
Although many aspects of active fluids have been studied, the effects of chirality on them are much less known. 
Here we study by computer simulation the dynamics of an unstructured droplet of chiral active fluid in three dimensions. 
Our model only considers the simplest possible combination of chiral and achiral active stresses, yet this leads to an unprecedented range of complex motilities,
including oscillatory swimming, helical swimming, and run-and-tumble motion.
Strikingly, while the chirality of helical swimming is the same as the microscopic chirality of torque dipoles in one regime, 
the two are opposite in another.  
Some of the features of these motility modes resemble those of some single-celled protozoa, 
suggesting that underlying mechanisms may be shared by some biological systems and synthetic active droplets.
\end{abstract}

\maketitle
Non-equilibrium fluids in statistical physics fall into two main categories: they may be externally driven or active (or possibly, both).
In driven fluids, the energy is injected continuously into the system globally or at boundaries, \emph{e.g.} by steady shearing. On the other hand in active fluids, the energy is injected into the system locally by the constituent particles themselves~\cite{marchetti13}. (This energy is then subsequently dissipated as heat.) Many examples of active fluids are biological in nature, such as bacterial suspensions~\cite{zhou13}, tissues~\cite{angelini11} and actomyosin networks in the cytoskeleton of eukaryotic cells~\cite{kruse05}. 

In the case of bacterial suspensions, the flagella of the bacteria continuously stir the solvent, driving the system out-of-equilibrium, and also yielding self-propulsion.
In the case of actomyosin, the myosin motors pull the actin filaments together causing them to contract lengthwise. 
Actomyosin contraction, in particular, is implicated in the swimming motility in MDCK tumour cells~\cite{poincloux10} and in some single-celled micro-organisms~\cite{ward14,kan14}.
In addition to these local sources of energy injection, many such fluids also have microstructures that show liquid-crystal-like ordering~\cite{zhou13}, albeit with polar rather than nematic order~\cite{kruse05}.
Importantly, many of these active polar liquid crystals are also microscopically chiral~\cite{julicher14,zhou13}. For instance, the actin filaments which make up the actomyosin network are twisted into right-handed helices, while the flagella in bacterial self-propulsion are left-handed helices. Although many aspects of active fluids have been extensively studied, much less is known about the effects of chirality, particularly under confinement such as within a droplet.

Droplets of active fluid have been shown to display interesting swimming and crawling motility modes~\cite{dogic12,dogic14,tjhung12,tjhung15}, which resemble those of biological cells.
Our aim here is to study the effects of chirality on the swimming motility of an active droplet in three dimensions.
In particular we show that a droplet of chiral active fluid can display complex swimming dynamics including screw-like, run-and-tumble, helical and oscillatory motion: 
these new motility modes emerge as a consequence of chirality, and of the coupling between active torque and force dipoles.

\begin{figure}
\centering
\includegraphics[width=1.0\columnwidth]{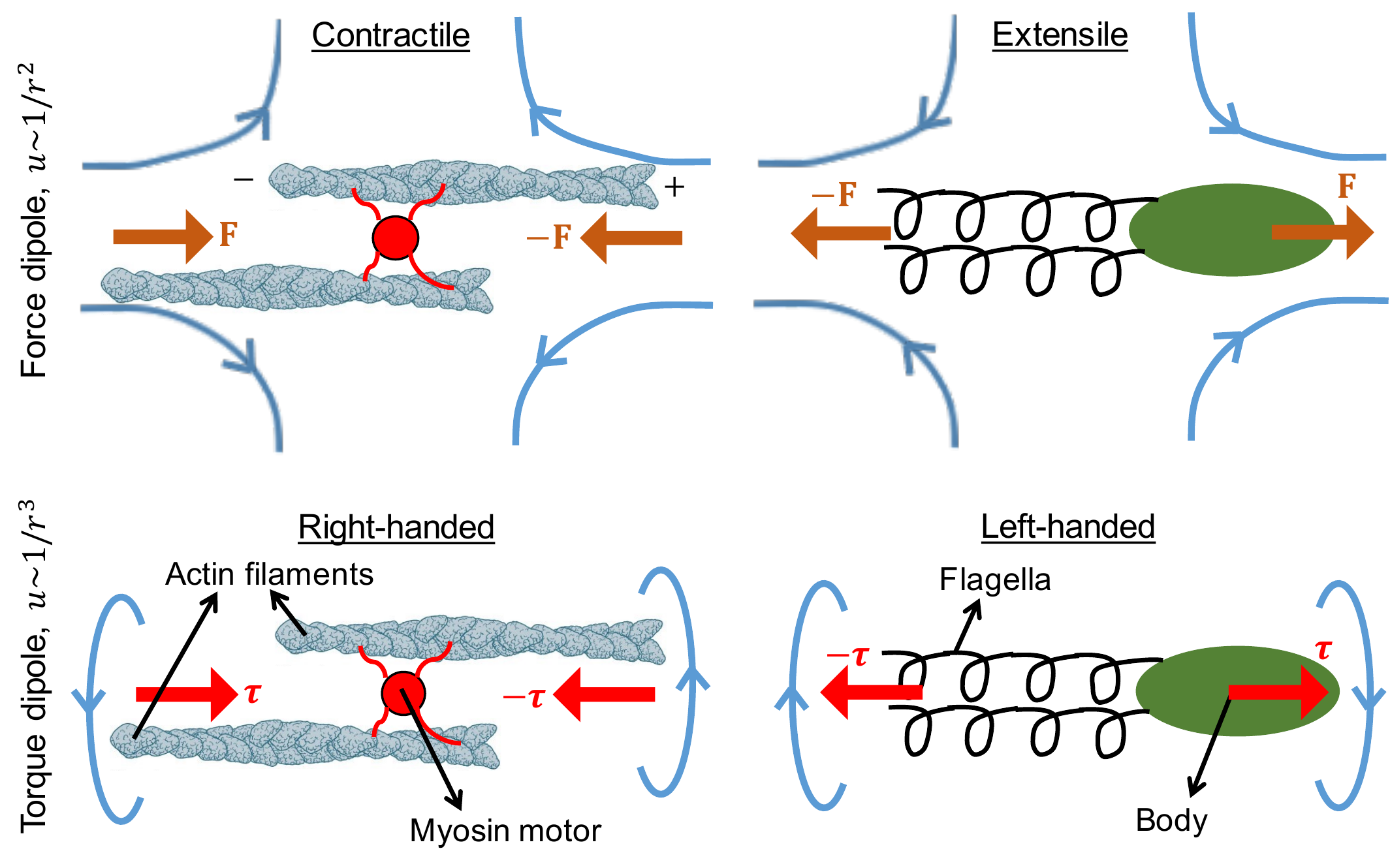}
\caption{
Left column: Actomyosin contraction inside the eukaryotic cells generates a force $(\mathbf{F},-\mathbf{F})$ and a torque dipole $(\mathbf{\tau},-\mathbf{\tau})$. (Only two actins cross-linked by a single myosin motor are shown.)
Right column: Flagella rotation in bacterial suspensions also generates a force $(-\mathbf{F},\mathbf{F})$ and a torque dipole $(-\mathbf{\tau},\mathbf{\tau})$, but in opposite directions. (Only a single bacterium is shown.)
The blue arrows in the figure indicates the fluid flows. 
Image of actin provided courtesy of the Mechanobiology Institute, National University of Singapore~\cite{actin}.}
\label{fig:1}
\end{figure}

Intriguingly, the helical swimming mode displayed by our simple model may be relevant to explain some features of the motility of single-celled protozoans such as \emph{Toxoplasma gondii}~\cite{ward14} and \emph{Plasmodium} ookinete~\cite{kan14}. 
(Toxoplasmosis is a common infection in cats and immunodeficient humans whereas \emph{Plasmodium} is the parasite responsible for malaria.) 
When measured \emph{in vitro} and \emph{in vivo}, the trajectories of these organisms in 3D almost invariably follow a left-handed helical path.
Yet the actin filaments are right-handed helices, hence, as explained in detail later on (Fig.~\ref{fig:1}), the natural microscopic chirality of actomyosin (thought to be responsible for the motility itself) is also right-handed. Thus there seems to be a mismatch between the microscopic and the observed macroscopic chiralities. 
One suggested explanation for this mismatch is that the actin filaments are somehow oriented along microtubules, which are left-handed~\cite{chiappino87}. 
We show here however that, in general, chirality selection in macroscopic motion need not involve any contest between chiralities at the molecular or local scale. 
Instead, at least within our model torque dipole system,
the competition between local twisting (chiral) and contraction (achiral) can reverse the handedness of the macroscopic motion without the presence of a second chiral component.
The mechanism of chirality reversal involves the polarization pattern of the active fluid at the droplet scale; this influences droplet shape and locomotion, and in turn depends on the competition between contractile and twisting forces locally. Note that this mechanism is different from spontaneous rotational or chiral symmetry breaking in active droplets with extensile forces and/or anchoring conditions~\cite{tjhung12,alexander16,maass16}, in diffusiophoretic nematic droplets~\cite{maass16}, or in swimmer aggregates~\cite{julius16,valeriani12}.
In all those cases, the droplets or the aggregates lack any intrinsic chirality, but start rotating following spontaneous breaking of chiral symmetry, so that they can rotate clockwise or anti-clockwise with equal probabilities.

\section*{Model and simulations}

We consider a droplet of active polar liquid crystal immersed in a passive isotropic fluid. The dynamical variables are a conserved scalar composition field $\phi(\mathbf{r},t)$, the polarization field $\mathbf{p}(\mathbf{r},t)$, and the fluid velocity field $\mathbf{u}(\mathbf{r},t)$.
Phenomenologically, the scalar field $\phi$ represents the concentration of active particles such as bacteria (in a swarm) or actin filaments (in the cytoskeleton). The polarization field $\mathbf{p}$ then represents the average orientation of these particles.
Finally $\mathbf{u}$ is the local average fluid velocity of active particles and solvent.
 
Our model includes two sources of active stress, arising from force dipoles and torque dipoles respectively. These force and torque dipoles arise naturally in the biological examples given above.
In the case of actomyosin contraction, the myosin motor pulls two actin filaments together, creating a quadrupolar fluid flow (see blue arrows in Fig.~\ref{fig:1} top left), which can be represented as a contractile force dipole. The filaments have a plus (or polymerizing) and a minus (or depolymerizing) end and they are aligned on average in the same direction as $\mathbf{p}$. Also, since the actin filaments are twisted in a right-handed direction~\cite{rice65}, the myosin motor tends to rotate them as it pulls, creating a pair of counter-rotating vortex flows (see blue arrows in Fig.~\ref{fig:1} bottom left), which we approximate as a right-handed torque dipole. Note that the helicity of a fiber is reversed by reflection but not physical rotation. Hence the torque dipole is unchanged if the polarization is reversed, and indeed would still be present if only one of the filaments was flipped over, creating a state of zero net polarization locally.

In the case of bacterial suspensions \emph{e.g. E. coli}, the activity is provided by the rotation of the flagella. The flagella rotate anti-clockwise whereas the body rotates clockwise resulting in a pair of counter-rotating vortex flows (see blue arrows in Fig.~\ref{fig:1} bottom right). Since the flagella are also helical, their rotation expels the fluid away from the bacterium, resulting in an extensile quadrupolar fluid flow (see blue arrows in Fig.~\ref{fig:1} top right). These two processes can then be approximated as an extensile force dipole and a left-handed torque dipole. Note that the directions of the force and the torque dipoles in bacteria are opposite to those in actomyosin contraction.

In our generic model, we therefore assume that a force and a torque dipole are embedded in each active particle with directions parallel or anti-parallel to $\mathbf{p}$. While we could also consider cases where the force dipole lies parallel and the torque dipole perpendicular to $\mathbf{p}$, and \emph{vice versa}, we are not aware of biological examples of this situation.
Since the force and the torque dipoles have a head-tail symmetry (but not the active particle itself), all the equations of motion below are symmetric under a global reversal of polarity: $\mathbf{p}(\mathbf{r},t)\rightarrow-\mathbf{p}(\mathbf{r},t)$, for all $\mathbf{r}$. There is an exception to this. If the active particles are swimming (as in the case of bacteria) or if there is an overall polymerization in one direction (as in the case of actomyosin), the global $\mathbf{p}\rightarrow-\mathbf{p}$ symmetry will be broken by a so-called `self-advection' term. (This causes the polarization pattern to move with fixed speed along $\mathbf{p}$ itself.)
Although it can be easily incorporated in the model~\cite{tjhung15}, for simplicity we do not consider self-advection in this paper. The same applies to anchoring terms that can also break $\mathbf{p}\leftrightarrow-\mathbf{p}$ symmetry~\cite{tjhung12}.  Even without such complications, we find a very rich dynamic phase diagram once local chirality is allowed for.

We consider a droplet of active polar fluid, defined such that $\phi\simeq\left|\mathbf{p}\right|\simeq1$ (active and polar) inside the droplet and $\phi\simeq\left|\mathbf{p}\right|\simeq0$ (passive and isotropic) outside the droplet.
This is achieved by introducing an equilibrium free energy (analogous to that of binary fluid) which stabilizes a phase coexistence between an active polar and a passive isotropic phase~\cite{tjhung12}. The equation of motion for the variable $\phi$ is then simply a conserved convective-diffusion equation whereas that for $\mathbf{p}$ is similar to the Leslie-Ericksen equation in liquid crystals~\cite{degennes-prost}.

Finally the fluid velocity $\mathbf{u}$ is described by the Stokes equation, basically a momentum balance equation:
\begin{equation}
0=-\nabla P + \eta\nabla^2\mathbf{u} + \nabla\cdot\underline{\underline{\sigma}}^p + \nabla\cdot\underline{\underline{\sigma}}^a
\end{equation}
where $P$ is the isotropic pressure, $\eta$ is the fluid viscosity and $\underline{\underline{\sigma}}^p$ is the passive elastic stress, which can be expressed as a combination of derivatives of the free energy.
The explicit form for $\underline{\underline{\sigma}}^p$ is analogous to that of standard equilibrium liquid crystals~\cite{degennes-prost} (see also Materials and Methods).
However as we have seen from Fig.~\ref{fig:1}, the force and the torque dipole from each active particle also pump the fluid solvent around it (resulting in quadrupolar fluid flow in the case of force dipoles or a pair of counter-rotating flows in the case of torque dipoles).
Therefore, there exists an additional non-equilibrium stress term in the Stokes equation which cannot be written as a derivative of the equilibrium free energy. This additional stress term is derived explicitly by coarse-graining the force and the torque dipoles~\cite{ramaswamy04,julicher12}, and the resulting expression is:
\begin{equation}
\sigma^a_{\alpha\beta} = \underbrace{\tilde{\zeta}_1 \phi p_\alpha p_\beta}_{\text{force dipoles}}   \\
		       + \underbrace{\tilde{\zeta}_2 \epsilon_{\alpha\beta\mu}\partial_\nu(\phi p_\mu p_\nu)}_{\text{torque dipoles}}
		       + \ldots   		
\label{eq:active-stress}		   
\end{equation}
where the Greek indices indicate Cartesian coordinates, $\epsilon_{\alpha\beta\gamma}$ is the Levi-Civita symbol and $\partial_\alpha$ is the partial derivative with respect to $r_\alpha$. Here $\tilde\zeta_{1,2}$ are activity parameters that quantify the force dipoles and torque dipoles respectively. 
The sign of $\tilde\zeta_1$ determines whether the force dipoles are contractile as in actomyosin or extensile as in bacteria (see Fig.~\ref{fig:1}); 
we shall focus only on contractile active stress, for which $\tilde\zeta_1>0$. 

Only the $\tilde\zeta_2$-term in Eq.~\ref{eq:active-stress} (which comes from the torque dipoles) is chiral; its divergence is the force density associated with an ensemble of microscopic torque dipoles. A different way of introducing chirality in our model is by adding a passive cholesteric term to the free energy; we do not discuss this here. In both cases, the sign of the chiral term determines whether the system is right- or left-handed at the local scale defined by the equations of motion. Therefore the chiral symmetry breaking is not spontaneous here, but imposed. For $\tilde\zeta_2>0$ (right-handed), the fluid flow from the torque dipole is as shown on the bottom left of Fig.~\ref{fig:1}; this is similar to the motion of closing a bottle cap. On the other hand for $\tilde\zeta_2<0$ (left-handed), the fluid flow is reversed (see the bottom right of Fig.~\ref{fig:1}) and hence resembles opening a bottle cap.
We define the torque dipole on the bottom left of Fig.~\ref{fig:1} to be right handed because if we imagine applying such a pair of torques on both ends of a piece of rope, it will tend to twist the rope into a right-handed helix.
With this convention, the activity of actomyosin contraction in eukaryotic cells is right-handed whereas that of bacterial suspension is left-handed. In this paper it is sufficient to focus only on right-handed activity $\tilde\zeta_2>0$ since the result for $\tilde\zeta_2<0$ is just its mirror image.

The forms for the active stress in Eq.~\ref{eq:active-stress} is rather general and can be obtained without any knowledge of actomyosin contraction in the cytoskeleton or flagellar rotation in bacteria.
Indeed, these follow from a hydrodynamic multipole expansion~\cite{adhikari14,lauga14} so long as the net force and the net torque are both zero. (Otherwise, the system is no longer active but a driven system like a colloid being pulled by an optical tweezer or gravity.)
In such an expansion, the lowest order contribution to the active stress comes from the force dipole whereas the lowest order chiral contribution comes from the torque dipole which is effectively a force quadrupole (since a point torque can be decomposed into a pair of forces). Higher order terms come from other force quadrupoles, six-poles, eight-poles, \emph{etc}.
The far-field flow from the force dipole scales with distance $r$ as $u\sim1/r^2$, whereas that of the force quadrupoles is one order higher: $u\sim1/r^3$. There are three independent types of force quadrupole~\cite{adhikari14,lauga14} of which the torque dipole is the only chiral one. By retaining only this term, our model isolates the new physics introduced by leading-order chirality.

To characterize the motion of the droplet, we first define the centre of mass (CM) velocity of the droplet to be:
$\tilde{\mathbf{V}}(t) = \int d\mathbf{r}\,\phi(\mathbf{r},t)\mathbf{u}(\mathbf{r},t) / \int d\mathbf{r}\,\phi(\mathbf{r},t)$
where the integration range is over the droplet phase (for all $\mathbf{r}$ such that $\phi(\mathbf{r},t)>0.5$). 
We then also define the angular velocity of the droplet about its CM to be:
$\tilde{\mathbf{\Omega}}(t) = \int d\mathbf{r}\,\phi \frac{\Delta\mathbf{r}\times\Delta\mathbf{u}}{\left| \Delta\mathbf{r} \right|^2} / \int d\mathbf{r}\,\phi$
where $\Delta\mathbf{r}=\mathbf{r}-\mathbf{R}$ with $\mathbf{R}(t)$ the CM of the droplet and $\Delta\mathbf{u}=\mathbf{u}-\mathbf{V}$. The integration range is again over the droplet phase. 

We perform 3D hydrodynamic simulations using a hybrid lattice Boltzmann method~\cite{stratford09,denniston04}. All the results below are presented in dimensionless units:
$\mathbf{V}=\tilde{\mathbf{V}}\eta R/\kappa$, 
$\mathbf{\Omega}=\tilde{\mathbf{\Omega}}\eta R^2/\kappa$, 
$\zeta_1=\tilde{\zeta}_1R^2/\kappa$ and 
$\zeta_2=\tilde{\zeta}_2R/\kappa$. 
($R$ is the radius of the droplet and $\kappa$ is the elastic constant of the active gel; for more details see Materials and Methods.)

\section*{Results}

\begin{figure}[tbhp]
\centering
\includegraphics[width=1.0\columnwidth]{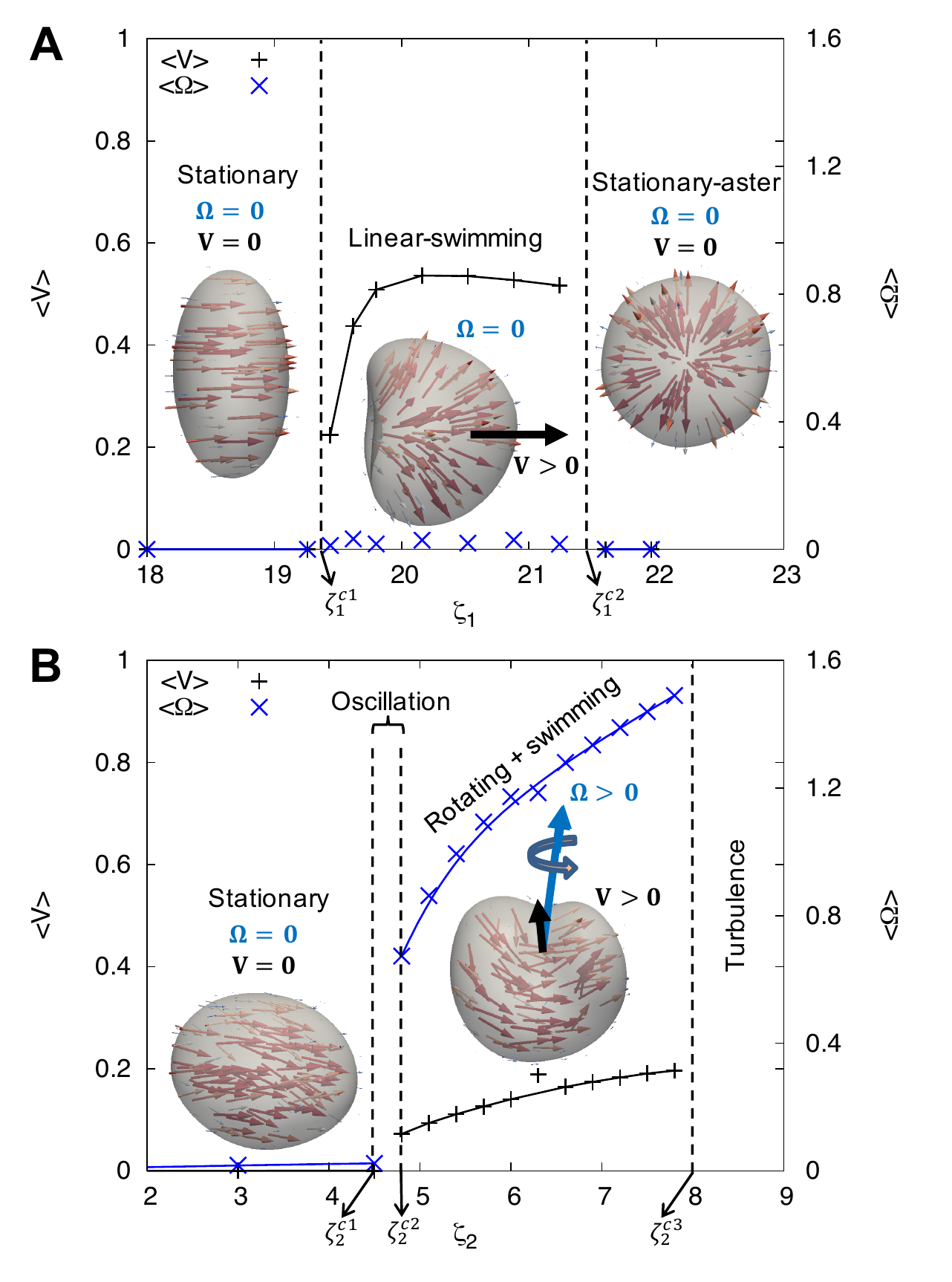}
\caption{
(A) Phase diagram for an active droplet with only force dipole activity ($\zeta_1>0$, $\zeta_2=0$). The plot shows the averaged centre of mass speed $\left< V \right>$ and angular speed $\left< \Omega \right>$ as a function of force dipole activity $\zeta_1$. Insets show the snapshots of the droplet morphology (grey surface) and the corresponding polarization field $\mathbf{p}$ (red arrows) for the three different phases. Numerical precision is $\pm0.01$ for $V$ and $\Omega$.
(B) Similar phase diagram for an active droplet with only torque dipole activity ($\zeta_1=0$, $\zeta_2>0$). (All quantities are in dimensionless units, see main text.)}
\label{fig:2}
\end{figure}

\subsection*{Force dipoles only}

First, we recapitulate the behavior of contractile droplets without chirality, \emph{i.e.,} $\zeta_1>0$ and $\zeta_2=0$. This case has been considered previously as a minimal model for cell swimming in 3D matrigel~\cite{hawkins11,tjhung12} and extended to cell crawling on a 2D surface~\cite{levine10,aranson13,tjhung15}. Fig.~\ref{fig:2}A shows the magnitudes of the CM velocity $\left<V\right>$ and angular velocity $\left<\Omega\right>$ of the droplet as a function of force-dipole activity $\zeta_1$. (The angle brackets indicate time-averaging at steady state.) For small values of activity $0<\zeta_1<\zeta_1^{c1}$, the droplet only slightly contracts in the direction of $\mathbf{p}$ (red arrows) without any translational or rotational motion ($\left<V\right>=\left<\Omega\right>=0$, stationary phase). 

At higher values of activity $\zeta_1^{c1}<\zeta_1<\zeta_1^{c2}$, the droplet spontaneously swims in a straight line without any rotation ($\left<V\right>>0$ and $\left<\Omega\right>\simeq0$, linear swimming phase). This spontaneous swimming is mediated by a splay instability in the polarization field $\mathbf{p}$ which creates an asymmetric flow field inside and outside the droplet~\cite{tjhung12}.
The (continuous) transition between quiescent and self-motile droplet occurs as the dimensionless parameter $\zeta_1=\tilde{\zeta}_1 R^2/\kappa$, measuring the ratio between active dipolar force and elasticity, exceeds a critical threshold (see SI).

Finally at even larger values of activity $\zeta_1^{c2}<\zeta_1$, the contractile activity can stabilize an aster defect in the polarization field $\mathbf{p}$ and the droplet becomes stationary again ($\left<V\right>=\left<\Omega\right>=0$, stationary-aster phase).

\subsection*{Torque dipoles only}
We next examine the effects of the torque dipole term $\propto\zeta_2$ in Eq.~\ref{eq:active-stress}. For simplicity, we first consider a droplet of active fluid with only torque dipoles, and no force dipole contribution ($\zeta_1=0$, $\zeta_2>0$). 

\begin{figure}
\centering
\includegraphics[width=1.0\columnwidth]{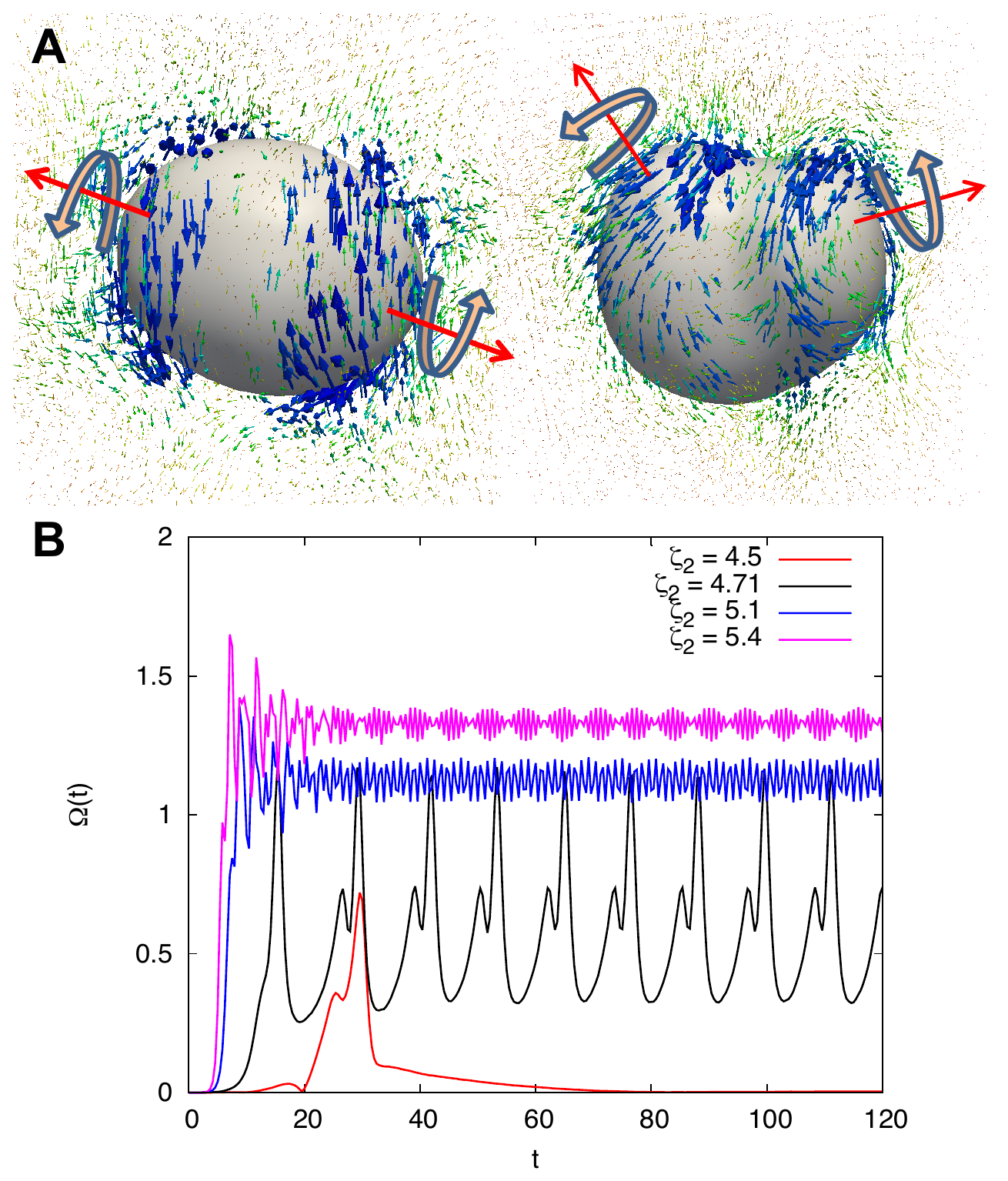}
\caption{
(A) The fluid velocity $\mathbf{u}$ for the stationary phase (left) and the rotating phase (right) corresponding to left and middle pictures in Fig.~\ref{fig:2}B.
(B) shows the time evolution of the angular speed $\Omega(t)$ for different values of torque dipole activity $\zeta_2$ (while $\zeta_1=0$). At some intermediate value $\zeta_2=4.71$, $\Omega(t)$ oscillates at steady state (see Suppl. Mov. 2). 
(The residual small oscillations at $\zeta_1=5.1$ and $\zeta_2=5.7$ are likely due to wobbles in the moving droplet.)}
\label{fig:3}
\end{figure}

Fig.~\ref{fig:2}B shows the averaged CM speed $\left<V\right>$ and angular speed $\left<\Omega\right>$ as a function of torque dipole activity $\zeta_2$ at steady state. For small values of activity $0<\zeta_2<\zeta_2^{c1}$, the droplet remains stationary ($\left<V\right>=\left<\Omega\right>=0$, stationary phase). The polarization field $\mathbf{p}$ inside the droplet, in this stationary state, is slightly twisted (through about a quarter of a helical pitch). Interestingly, the flow field $\mathbf{u}$ outside the droplet forms a pair of counter-rotating vortices (see Fig.~\ref{fig:3}A left). Thus the droplet behaves coherently as a single, right-handed torque dipole. Similar counter-rotating flows have also been observed in \emph{C. elegans} embryos~\cite{julicher14}, possibly indicating that eukaryotic cells have intrinsic chirality.

At higher values of the chiral activity, $\zeta_2^{c2}<\zeta_2<\zeta_2^{c3}$, the droplet acquires both translation and rotation ($\left<V\right>>0$ and $\left<\Omega\right>>0$, rotating phase).
This rotation is mediated by a bend-instability in $\mathbf{p}$ (in contrast with the splay produced by force dipoles) which causes an asymmetric, counter-rotating flow field $\mathbf{u}$ (see Fig.~\ref{fig:3}A right). Meanwhile, the interface of the droplet also follows the direction of $\mathbf{p}$, forming a chiral bent shape. 
The direction of rotation $\mathbf{\Omega}$ aligns with that of bending $(\nabla\times\mathbf{p})\times\mathbf{p}$, \emph{i.e.,} perpendicular to the overall polarization $\mathbf{p}$.
Also note that in the rotating phase, the droplet's swimming velocity $\mathbf{V}$ points roughly parallel to $\mathbf{\Omega}$. Qualitatively, the motion of the droplet is like the motion of a right-handed screw going into the wall (see Supplementary Movie 1). (It is also like the motion when the screw is extracted from the wall, since both $\mathbf{V}$ and $\mathbf{\Omega}$ are then reversed.) Here, therefore the chirality of the trajectory is determined by the underlying microscopic chirality. 

There also exists a small window at $\zeta_2^{c1}<\zeta_2<\zeta_2^{c2}$ where the droplet oscillates between the rotating and the stationary phase (oscillatory phase), 
corresponding to a subcritical Hopf bifurcation in dynamical systems (see Supplementary Figure 1). 
If we look at the time evolution of the CM angular speed $\Omega(t)$ in this window (\emph{e.g.} $\zeta_2=4.71$ in Fig.~\ref{fig:3}B), 
$\Omega(t)$ oscillates between some large values (rotating) and small values (approximately stationary). 
Qualitatively, the droplet remains stationary for some time then suddenly rotates for a while, then becomes stationary again and so on (see Supplementary Movie 2).

Finally, at much higher values of activity $\zeta_2^{c3}<\zeta_2$, the fluid flow inside and outside the droplet becomes turbulent with large interfacial fluctuations (turbulent phase in Fig.~\ref{fig:2}A).

The transition between quiescent and spontaneously rotating droplet is governed by the dimensionless parameter $\zeta_2=\tilde{\zeta}_2 R/\kappa$, 
which measures the ratio between active forces, due to the torque dipole, and elastic forces. 
As a result, the critical torque threshold $\tilde{\zeta}_2^{c1}$ scales with $R$ differently with respect to the critical force dipole discussed in the previous Section. 
A distinct important quantity is the capillary number $\text{Ca}_2=\tilde{\zeta}_2/\gamma$, where $\gamma$ is the surface tension of the droplet: 
when this number is large, active torques lead to a change in droplet shape.
The analogous capillary number relevant for force dipoles is $\text{Ca}_1=\tilde{\zeta}_1 R/\gamma$.

\subsection*{Force and torque dipoles combined}
\begin{figure}
\centering
\includegraphics[width=1.0\columnwidth]{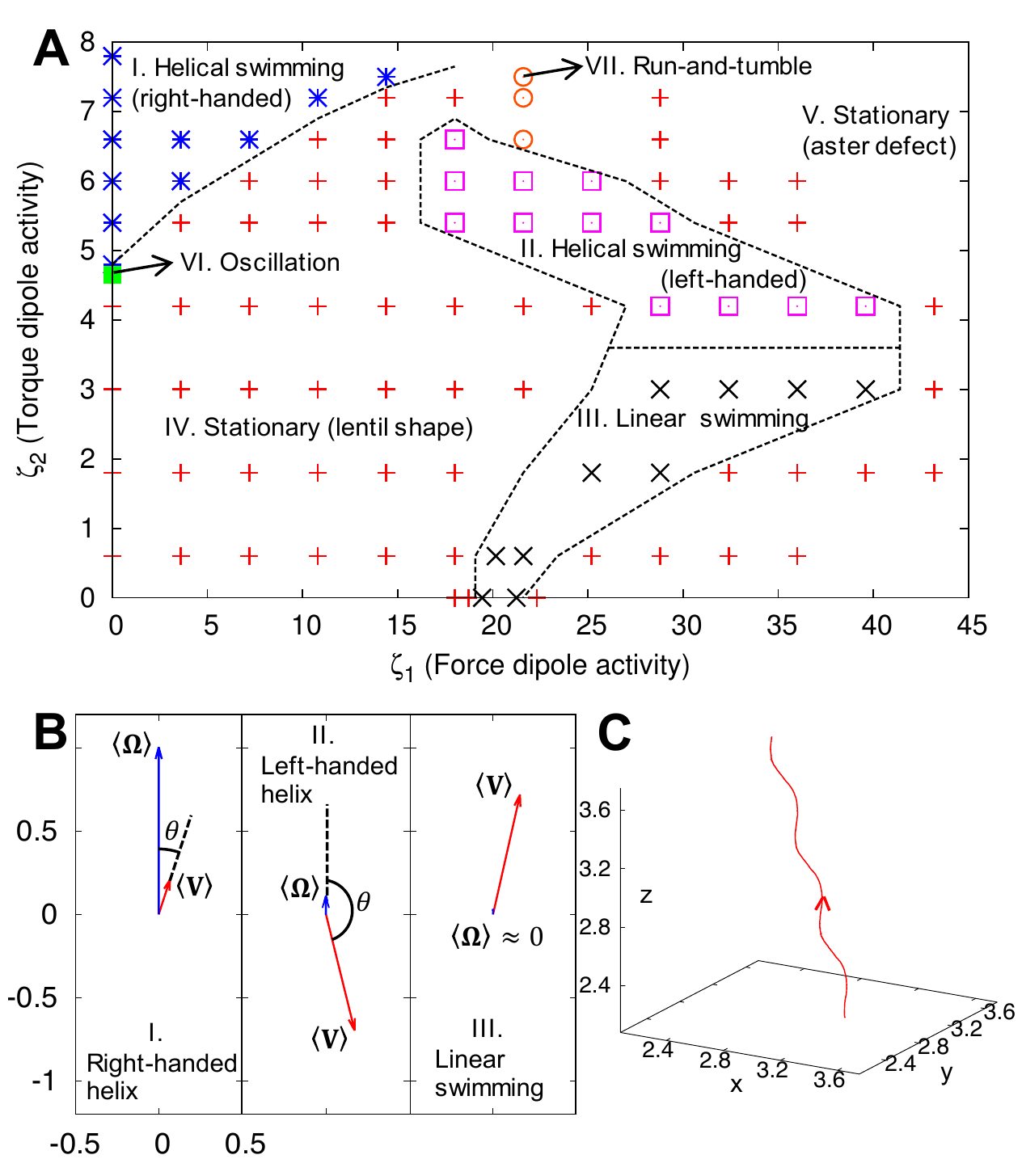}
\caption{
(A) Phase diagram in the full $\zeta_1$-$\zeta_2$  parameter space can be divided into several different phases.
Note that there are two helical swimming phases: left- (phase I) and right- (phase II) handed helix.
(B) shows the typical time-averaged CM velocity vector $\left< \mathbf{V} \right>$ and angular velocity vector $\left< \mathbf{\Omega} \right>$
for phase I (left panel), phase II (middle panel), and phase III (right panel).
(C) shows the typical right-handed helical trajectory corresponding to phase I 
(parameters used: $\zeta_1=10.8$ and $\zeta_2=7.8$, the lengths are in units of droplet's radius).}
\label{fig:4}
\end{figure}

Now we consider the effects of both force and torque dipoles being simultaneously present in an active droplet, which is the generic case. Fig.~\ref{fig:4}A shows the phase diagram in the $\zeta_1$-$\zeta_2$ parameter space  (assuming both $\zeta_1$ and $\zeta_2$ are positive as before).
From this phase diagram, we can distinguish several different dynamical regimes or `phases' for the behavior of the droplet at steady state. First, on the lower left hand corner, we have a stationary phase (labelled IV), where the droplet remains stationary without any translational or rotational motion. This droplet has a lentil shape similar to that depicted on the left picture of Fig.~\ref{fig:2} A (see also~\cite{tjhung12}). On the right hand side of the phase diagram, we have another stationary phase (labelled V) with an aster defect at the centre of the droplet, now similar to that depicted on the right picture of Fig.~\ref{fig:2}A. In these two limiting regimes, the chiral activity barely changes the phases seen for a purely contractile system.

Next, we observe two helical phases (labelled I and II) in the phase diagram. In these two phases, the droplet swims in a helical trajectory at steady state (see Fig.~\ref{fig:4}C for a typical helical trajectory). As announced earlier, we can have both a right-handed helix (phase I) and a left-handed helix (phase II) on the same phase diagram, even though the torque-dipole activity $\zeta_2$ is always positive or right handed. In the case of a right-handed helix (phase I), the angle between $\mathbf{V}$ and  $\mathbf{\Omega}$ is always less than $90^\circ$ (see Fig.~\ref{fig:4}B left panel). On the other hand for a left-handed helix (phase II), the angle between $\mathbf{V}$ and $\mathbf{\Omega}$ is always larger than $90^\circ$ (see Fig.~\ref{fig:4}B middle panel). These identifications swap over if the chiral activity parameter $\zeta_2$ is reversed in sign. Among these helical trajectories are screw-like ones in which $\mathbf{V}$ and $\mathbf{\Omega}$ are almost parallel, but also others where the angle between these is large.

\begin{figure}
\centering
\includegraphics[width=0.7\columnwidth]{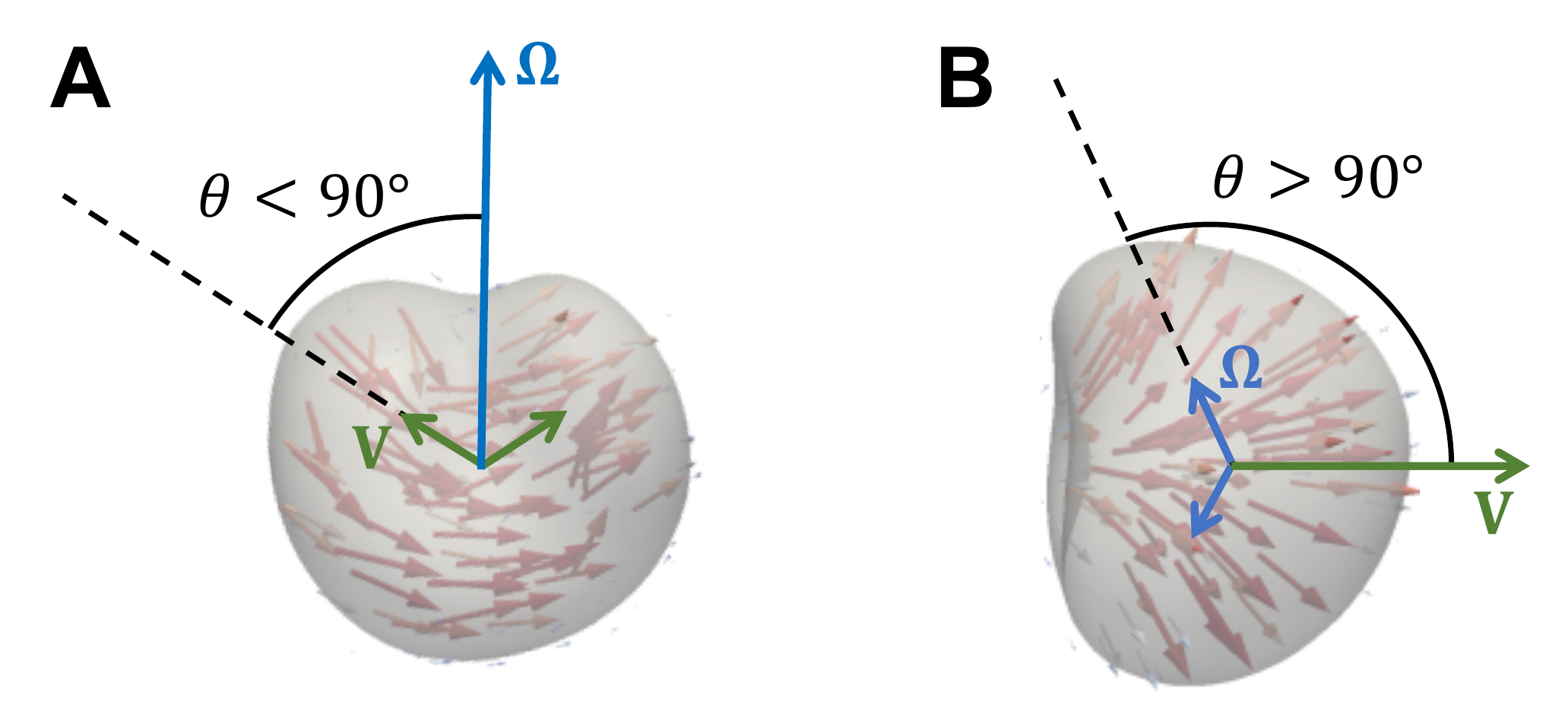}
\caption{
A purely right-handed microscopic chirality can give rise to both right-handed and left-handed macroscopic motion due to two different ways the polarization field $\mathbf{p}$ (red arrows) is arranged inside the droplet. (A) Right handed helix: bend deformation dominates. The green arrows indicate two possible directions for $\mathbf{V}$ (B) Left handed helix: splay deformation dominates. Similarly, the blue arrows indicate two choices of many possible directions for $\mathbf{\Omega}$.}
\label{fig:5}
\end{figure}

The presence of both right-handed and left-handed helical motion from a purely right-handed microscopic chirality can now be understood as follows. The right-handed helix (phase I) arises when the torque dipole is large and the force dipole weak. To explain it, consider adding a small force dipole perturbation to the purely torque dipolar rotating phase in the middle of Fig.~\ref{fig:2}B.
The contractile perturbation induces a translational velocity along $(\nabla.\mathbf{p})\mathbf{p}\sim\pm\mathbf{p}$ as stated previously; because of the pre-existing bending created by the chiral activity (see Fig.~\ref{fig:5}A), the resulting angle between $\mathbf{V}$ and $\mathbf{\Omega}$ will be less than $90^\circ$.
Conversely, for the left-handed helix (phase II), consider adding a small torque dipole perturbation to the purely contractile linear-swimming phase in the middle of Fig.~\ref{fig:2}A.
The torque dipole now induces an angular velocity perpendicular to the overall polarization $\mathbf{p}$ (see Fig.~\ref{fig:5}B).
Because of the pre-existing splay in the contractile swimming state, the resulting angle between $\mathbf{V}$ and $\mathbf{\Omega}$ will be larger than $90^\circ$.
We see that the left-right helix distinction is due to having very different configurations of the polarization field $\mathbf{p}$ inside the droplet, depending on the outcome of a competition between chiral and contractile activity.
In the case of phase I (large chirality) bend deformation dominates, whereas in the case of phase II (large contractility), splay dominates.
Consequently phase I droplets also have quite distinct morphology from phase II droplets, 
with chiral bent shape for the former and a dimple for the latter (see Fig.~\ref{fig:5}A,B respectively).

We emphasise again that our study addresses the effects of nonzero chirality at microscopic scales on macroscopic droplet trajectories. 
Similar, macroscopically helical motions can also be obtained via spontaneous symmetry breaking (SSB) from zero microscopic chirality.
The two cases are easily distinguished, however, since in the SSB scenario each swimmer chooses between left and right helicity at random; both states exist at the same point in the phase diagram. 
In our case, left and right macroscopic helicity are located at different state points.

In the phase diagram Fig.~\ref{fig:4}A, we also have a linear-swimming phase (labelled III), where the droplet swims in a nearly-straight line, similar to the picture in the middle of Fig.~\ref{fig:2}.
In this phase, $\mathbf{\Omega}$ is approximately zero (within numerical precision $\pm0.01$) while $\mathbf{V}$ is approximately constant at steady state (see Fig.~\ref{fig:4}B right panel); because of small residual helicity this is separated from phase II by a crossover rather than a sharp transition line. On the $\zeta_2$-axis in the phase diagram, we have the oscillatory phase (labelled VI), where $V(t)$ and $\Omega(t)$ oscillate at steady state as discussed above.
Intriguingly, we also discover a run-and-tumble phase (labelled VII), where the droplet is observed to swim in a straight line for some period of time (run) and then abruptly changes its direction (tumble) before swimming again in a straight line and so on (see Supp. Mov. 3).
Note that due to the absence of noise in our system, at each tumble phase, the droplet changes its direction at fixed angle. 
(This is similar to zig-zag motion in a particle model~\cite{tarama16}.)
The run-and-tumble motion, again, arises here as an example of oscillatory behaviour: it requires simultaneous presence of force and torque dipoles, as well as proximity to the transition between stationary and helical phase.

\section*{Discussion}
We have shown within a simplified model how the microscopic chirality of active particles (such as actin filaments or bacteria) can translate into a macroscopically chiral motion at the scale of a droplet. 
While other models for helical droplet swimming have been proposed based on particle dynamics (e.g.~\cite{hiraiwa09,yamamoto16}), the current one is the simplest to our knowledge based on active gel theory. 
In our framework, helical trajectories can arise for swimming droplets from a combination of contractile and chiral activities.
A right-handed microscopic chirality does not necessarily translate into a right-handed helical trajectory, because the macroscopic helicity is mainly controlled by the pattern of the polarization field $\mathbf{p}$ (describing the average alignment of particles) within the droplet (see Fig.~\ref{fig:5}). This configuration is determined by a contest between contractility and chirality, not between two different sources of chirality. While our study addresses an active gel droplet, the helical swimming mode we found may be relevant to understand some features of the 3D swimming of protozoans such as {\it Toxoplasma}, whose motility mechanism is still elusive~\cite{egarter14}. Notably, such organisms often display a mismatch between microscopic and macroscopic chirality not unlike the one we discovered here.

Interestingly, we are also able to obtain an oscillatory state without any built-in dynamical oscillator in our model. This oscillatory behaviour is also observed in simple models of cell crawling~\cite{tjhung15} and actin cortex~\cite{whitfield16}. This illustrates how simple physical mechanisms can give rise to oscillatory behaviour in biological systems without the need of an internal chemical oscillator.

In conclusion, hydrodynamic instabilities in a droplet of generic chiral active fluids can lead to a striking variety of swimming motilities in 3D. 
It would be of interest in the future to search for biological analogues of these mechanisms, which rely on the coupling between orientational order and active torques and forces. 
This work has focussed on the effects of chirality on swimming through a bulk medium rather than crawling across a solid surface. 
Since molecular chirality is present at some level in all cell types and micro-organisms, it would be very interesting to look for its signatures in crawling processes also; we hope to return to this in future work.




\section*{Materials and methods}
We introduce a free energy functional $F[\phi,\mathbf{p}]$ to stabilize a droplet of active polar phase ($\phi\simeq\left|\mathbf{p}\right|\simeq1$) 
in the bulk of a passive isotropic phase ($\phi\simeq\left|\mathbf{p}\right|\simeq0$):
\begin{eqnarray}
F[\phi,\mathbf{p}] &=& \int d\mathbf{r}\, \left\{ a\phi^2(\phi-1)^2 + \frac{k}{2}\left|\nabla\phi\right|^2 \right. \nonumber \\ 
			         &-& \left. \alpha\left(\phi-\frac{1}{2}\right)\left|\mathbf{p}\right|^2 + \frac{\alpha}{4}\left|\mathbf{p}\right|^4 + \frac{\kappa}{2}\left(\nabla\mathbf{p}\right)^2 \right\}
\end{eqnarray}
where $a$, $k$, $\alpha$ and $\kappa$ are positive constants. 
$a$ and $k$ are related to the surface tension $\gamma$ and interfacial width $\ell$ via: $\gamma=\frac{8}{3}\sqrt{ka}$ and $\ell=\sqrt{\frac{k}{2a}}$.
$\kappa$ is the (single, for simplicity) elastic constant controlling bend, twist and splay deformation in $\mathbf{p}$.

The dynamics of $\phi(\mathbf{r},t)$ follows a conserved diffusion-reaction equation:
\begin{equation}
\frac{\partial\phi}{\partial t} + \nabla\cdot\left( \phi\mathbf{u} - M\nabla\frac{\delta F}{\delta\phi} \right) = 0 
\end{equation}
where $M$ is the mobility. The dynamics of $\mathbf{p}(\mathbf{r},t)$ is similar to the Leslie-Ericksen equation for liquid crystals~\cite{degennes-prost}:
\begin{equation}
\frac{\partial\mathbf{p}}{\partial t} + \mathbf{u}\cdot\nabla\mathbf{p} = 
							- \underline{\underline{\Omega}}\cdot\mathbf{p}
							+ \xi\underline{\underline{v}}\cdot\mathbf{p}
							- \frac{1}{\Gamma}\frac{\delta F}{\delta\mathbf{p}}
\end{equation}
where $\underline{\underline{v}}$ and $\underline{\underline{\Omega}}$ are the symmetric and anti-symmetric part of the velocity gradient tensor $\partial_\alpha u_\beta$ respectively. Here $\xi$ is the shear aligning parameter ($\xi>1$ for shear-aligning rod-shaped particles) and $\Gamma$ is the relaxation timescale for $\mathbf{p}$. Finally the equation for $\mathbf{u}(\mathbf{r},t)$ is described by the Stokes equation (see main text). The form of the passive elastic stress $\underline{\underline{\sigma}}^p$ in the Stokes equation is again similar to nematic liquid crystals (with $\mathbf{h}=-\frac{\delta F}{\delta\mathbf{p}}$)~\cite{degennes-prost}:
\begin{equation}
\sigma^p_{\alpha\beta} = \frac{1}{2}(p_\alpha h_\beta - p_\beta h_\alpha) - \frac{\xi}{2}(p_\alpha h_\beta + p_\beta h_\alpha) 
					- \kappa\partial_\alpha p_\gamma\partial_\beta p_\gamma
\end{equation}
These equations have been used in many other studies~\cite{marchetti13,kruse05,zhou13}.

\begin{acknowledgements}
We thank R. Adhikari, E. Lauga, C.~A. Whitfield, and G.~E.~Ward for illuminating discussions. MEC holds a Royal Society Research Professorship.
\end{acknowledgements}


\end{document}